\documentclass[aps,pra,reprint,superscriptaddress]{revtex4-2}
\pdfoutput=1

\usepackage[margin=0.9in, top=1.1in]{geometry}
\usepackage{amssymb}
\usepackage{dsfont}
\usepackage{complexity}
\usepackage[utf8]{inputenc} 
\usepackage[T1]{fontenc}    
\usepackage{hyperref}       
\usepackage{url}            
\usepackage{booktabs}       
\usepackage{amsfonts}       
\usepackage{nicefrac}       
\usepackage{microtype}      
\usepackage{lipsum}
\usepackage{bbm}
\usepackage{bm}
\usepackage{physics}
\usepackage{placeins}
\usepackage{bbold}
\usepackage{tikz}
\usepackage{forest}
\usepackage{multirow}
\usepackage{algorithm}
\usepackage{algpseudocode}
\usepackage{dsfont} 
\usetikzlibrary{calc,arrows.meta,positioning, shapes.geometric, arrows, shapes.symbols,shapes.callouts,patterns}
\usepackage{qcircuit}
\usepackage{graphicx}
\usepackage{graphics}
\usepackage{amsthm}
\usepackage[capitalise]{cleveref}
\usepackage[caption=false]{subfig}
\usepackage{latexsym}
\usepackage{mathtools}
\usepackage{verbatim}
\usepackage{upgreek}
\usepackage{textgreek}
\usepackage{thmtools}
\usepackage{thm-restate}
\usepackage{hyperref}

\newtheorem{theorem}{Theorem}
\newtheorem{definition}[theorem]{Definition}
\newtheorem*{remark}{Remark}

\newcommand{\x}{x}
\newcommand{\y}{y}
\newcommand{\z}{\mathbf{z}}
\newcommand{\rin}{\ensuremath{\mathtt{in}}}
\newcommand{\rval}{\ensuremath{\mathtt{val}}}

\newcommand{\subfigimg}[3][,]{%
  \setbox1=\hbox{\includegraphics[#1]{#3}}
  \leavevmode\rlap{\usebox1}
  \rlap{\hspace*{0.8\linewidth}\raisebox{\dimexpr\ht1-2\baselineskip}{#2}}
  \phantom{\usebox1}
}

\newcommand{\UTS}{\affiliation{%
Centre for Quantum Software and Information,
University of Technology Sydney, Sydney, NSW 2007, AU}}

\newcommand{\UMD}{\affiliation{%
Joint Center for Quantum Information and Computer Science, University of Maryland, College Park, MD 20742, USA}}     

\newcommand{\PKU}{\affiliation{%
Beijing International Center for Mathematical Research, Peking University, Beijing, China}}

\newcommand{\BBQP}{\affiliation{%
BosonQ Psi (BQP) Corporation, New York, USA}}

\begin{document}

\title{Assessing Quantum and Classical Approaches to Combinatorial Optimization: Testing Quadratic Speed-ups for Heuristic Algorithms}

\author{Pedro C. S. Costa} 
\email{pcs.costa@protonmail.com}\BBQP\UTS
\author{Mauro E.S. Morales}\UMD
\author{Dong An}\PKU\UMD
\author{Yuval R. Sanders}
\email{yuval.sanders@uts.edu.au}\UTS 

\begin{abstract}
Many recent investigations conclude, based on asymptotic complexity analyses,
that quantum computers could accelerate combinatorial optimization (CO) tasks relative to a purely classical computer.
However, asymptotic analysis alone cannot support a credible claim of quantum advantage.
Here, we highlight the challenges involved in benchmarking quantum and classical heuristics for combinatorial optimization (CO), with a focus on the Sherrington-Kirkpatrick problem.
Whereas hope remains that a quadratic quantum advantage is possible,
our numerical analysis casts doubt on the idea that current methods exhibit any quantum advantage at all.
This doubt arises because even a simple classical approach can match with quantum methods we investigated.
We conclude that more careful numerical investigations are needed to evaluate the potential for quantum advantage in CO,
and we give some possible future directions for such investigations.
\end{abstract}

\maketitle

There is compelling reason to believe that heuristic quantum optimization algorithms may achieve quadratic asymptotic complexity advantages
over classical competitors~\cite{Montanaro_2016, dalzell2023quantumalgorithmssurveyapplications}.
However, such asymptotic analyses eschew several important real-world considerations such as differing runtime costs for primitive operations on classical and quantum computers.
Whereas some recent work~\cite{sanders2020compilation} has attempted to draw a direct comparison between wall-clock runtimes for large quantum and classical computers,
such work has only increased doubts about the real-world value of quadratic quantum advantages~\cite{Babbush2021focusbeyondquadratic}.

One might object that such analyses are unreliable because they take an overly simplistic view about how to compare classical and quantum strategies.
We take this objection seriously.
Here we seek to overcome difficulties in the analysis from Ref.~\cite{sanders2020compilation}
by identifying and executing a strategy for directly and numerically comparing classical and quantum performance.
We are particularly interested in the extent to which an apparent quantum advantage may be a result of mistaken comparisons between quantum and classical algorithms.

\begin{figure}[t]
\centering
\includegraphics[width = 0.48\textwidth]{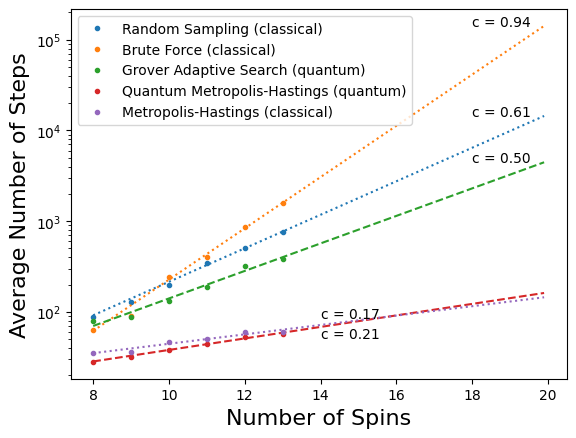}\\
    \caption{\textit{Performance of classical and quantum algorithms for solving the spin glass problem and the predicted asymptotic scaling $\mathcal{O}(N^c)$.}
    The abscissa depicts the number of spins in the spin chain and the ordinate shows the average number of steps (main subroutine calls associated with each technique used) made by the classical and quantum methods across one hundred randomly chosen problem instances.
    Each of the five methods is given its own color (online) and numerical results are indicated by
    a small solid circle.
    We have also plotted exponential fits of the form $b \times N^c$,
    and the plots are labelled with the fitted value of $c$ to indicate asymptotic scaling.
    We have used dashed lines for fits to quantum heuristics and dotted lines for classical
    to visually distinguish between classical and quantum strategies.
    }
    \label{fig:main_plot}
\end{figure}

To this end, we perform a numerical comparison between classical and quantum optimization heuristics.
We are careful to include both of what we call `structured' and `unstructured' approaches in each of the classical and quantum cases --
whereas a structured approach takes due note of any specific knowledge we have about mathematical properties of the objective function,
an unstructured approach would ignore such knowledge.
To avoid making this study needlessly complicated, the only structured approach we consider Metropolis-Hastings for simulated annealing.
This is a convenient choice because there is a quantum analog of classical Metropolis-Hastings~\cite{lemieux2020efficient} that we henceforth refer to as the LHPST walk after the authors' initials.
Our numerical analysis (\cref{fig:main_plot}) fails to reveal a performance difference between the classical and quantum Metropolis-Hastings approach.

There are well-documented difficulties in properly benchmarking even classical optimization algorithms~\cite{wolpert1997no, macready_wolpert}.
Specifically, Ref.~\cite{wolpert1997no} gives several No Free Lunch (NFL) theorems that address the question of how to select an appropriate black-box optimization strategy for a given problem.
The NFL theorems collectively suggest that we should not expect to find any one algorithm that is universally superior to all contenders;
in practice, one should expect at least to tune the hyper-parameters of a stochastic algorithm to particular problem sets.

It is not our intention to delve into all of these difficulties; that would be a big job.
Instead, we seek to showcase the impact of one benchmarking difficulty
on the comparison of classical and quantum optimization heuristics:
the impact of structured versus unstructured approaches to the search space.

To make this point, we digress briefly on the idea of treating an optimization problem as a search problem.
Supposing we have access to an oracle that determines whether a given candidate solution to the optimization problem is minimal or not,
we could take a ``brute force'' approach and test each candidate solution using the oracle until we find one that the oracle certifies is minimal.
Brute force  should not be considered a sensible approach in most practical situations
because we are usually aware of a significant amount of mathematical detail about our objective function,
which in turn should be expected to give us an indication of which candidate solutions are reasonable to test.
Nevertheless, in a purely unstructured search problem, one cannot do better than brute force.
Our knowledge about the mathematics of the search space modifies the optimization problem into a \emph{structured} search problem.

Grover’s algorithm is frequently proposed as a quantum deterministic method for solving search and optimization problems due to its theoretical capability to achieve quadratic speedup in unstructured search settings. The algorithm frames optimization as a black-box problem, aiming to identify either an element of a list or the optimal solution by querying a quantum oracle. 

One example of a Grover-inspired method  is the Grover Adaptive Search (GAS), which extends the iterative method of Grover's search \cite{Grover1996FastQuantum} in cases where the number of solutions is unknown \cite{tightbounds}. GAS has been explored in \cite{Durr1996minimum, baritompa2005grover, chakrabarty2017dynamic} and demonstrated a quadratic speedup over the brute force method for combinatorial optimization problems \cite{gilliam2021grover}. These Grover-inspired adaptations may approximate the oracle or utilize partial information about the solution space, however quadratic speedups over brute-force methods for optimization are not always guaranteed. This approach, while building on Grover’s framework, does not use any structure of the solution space like classical techniques such as Metropolis-Hastings and genetic algorithms.

Now that we have explained our goal of distinguishing between structured and unstructured
optimization heuristics, we give three key definitions.

\begin{definition}[MCOP]
\label{def:MCOP}
A \emph{minimization combinatorial optimization problem} (MCOP)
gives a real-valued objective function $f$ whose domain is a finite but large set.
The task is to identify an argument $x_{\rm opt}$ for the function $f$ such that
$f(x_{\rm opt})$ is globally minimal;
i.e. $f(x_{\rm opt}) \leq f(x)$ for any other $x$ in the domain of $f$.
An MCOP is said to have a \emph{polynomial-time approximation scheme} (PTAS)
if it has an $\epsilon$-approximation algorithm
(i.e.~the output $x$ of the algorithm satisfies $f(x) \leq (1-\epsilon) f(x_{\rm opt})$)
for every constant $\epsilon>0$, and the algorithm runs in polynomial time in the size of the instance for constant $\epsilon$.
\end{definition}

It is well-known that computing the approximated solution can be as difficult as computing the optimum solution for many problems~\cite{haastad1999clique},
ultimately resulting in a formal theorem about \emph{the hardness of approximation problems}.
The hardness of approximation is an important consequence of
the Probabilistically Checkable Proof (PCP) Theorems \cite{arora1998probabilistic}.
Via Ref.~\cite{arora1998proof}, we know that the PCP Theorem is equivalent to the statement
``MAX-E3SAT (NP-hard problem) has no PTAS assuming P $\neq$ NP'',
which proves the NP-hardness of approximation algorithms.

\begin{definition}[SKOP]
The \emph{Sherrington-Kirkpatrick optimization problem} (SKOP)~\cite{PhysRevLett.35.1792}
is an MCOP where the domain of the objective function $E$ is the set of length-$n$ strings
$\mathbf{s} = s_1 \dots s_n$ with $s_i = \pm 1$ and $E$ is quadratic in the sense that
$E(\mathbf{s}) = \sum_{i\ne j}^{n} w_{ij} s_i s_j$ for some given set of real numbers
$\left\{w_{ij}|i,j=1,\dots,n\right\}$ such that $w_{ij} = w_{ji}$.
\end{definition}

Although the focus of this paper is not to justify the practical applications of the Sherrington-Kirkpatrick (SK) Hamiltonian, we have in Ref.~\cite{kirkpatrick1983optimization} its clear application for the physical design of electronic circuits. The simplified version for designing electronic circuits, which can be represented by the SKOP, only considers the distribution of the electronic circuits over two connected chips; the weights $w_{ij}$ are the number of signals passing between circuits $i$ and $j$ with the two values $s_i=\{-1,1\}$ indicating which chip the circuit $i$ is placed. The goal is to find a configuration in which the signals propagating across the two chips --- which are slower than the signals inside the same chip --- are minimized.

For our purposes, there is a subtlety about the approximate solutions for the SK problem.
We are not trying to implement a PTAS but instead something simpler.

\begin{definition}[$\alpha$-AEAR]
Given $n$, an algorithm $\mathcal{A}$ for solving $n$-SKOP is an $\alpha$-\emph{average energy approximation ratio} ($\alpha$-AEAR) algorithm if whenever the ground state energy to an instance has value $E_{gs}<0$, $\mathcal{A}$ return a solution with at most $\alpha \cdot E_{gs}$ on the average energy
\begin{equation}
    E_{avg}\leq \alpha \cdot E_{gs}.
\end{equation}
\end{definition}
\begin{remark}
 The $E_{gs} < 0$ is met in all instances of the SK Hamiltonian analyzed in our benchmarking.
\end{remark}

An important difference of this work with those associated to the NFL theorem \cite{wolpert1997no, macready_wolpert} is that in previous work, only the number of cost function evaluations is compared among the algorithms. Here, in our benchmarking we compare the performance of both classical and quantum optimization algorithms over the same SKOP instances, keeping in mind that a single evaluation might require more clock time in the case of quantum algorithms. Related to the previous point, we do not fix the number of steps each algorithm performs and then compare the quality of the solution, but rather fix the quality of the solution and then compare the number of steps required to reach such solution and then estimate the clock time of each step. This provides a way to compare fairly both classical and quantum methods.

With these points in mind, we now turn to discuss the benchmarking methodology used when comparing the optimization algorithms.
For the classical methods, specifically Random Sampling (RS) and Metropolis-Hastings (MH), we evaluated each SKOP instance by determining the number of steps (main subroutine), \(T\) to achieve 0.9-AEAR. Likewise, for the LHPST we determine the \(T\) quantum walk steps until we get the 0.9-AEAR. In both cases classical and quantum we previously computed the ground state energy, $E_{gs}$, of the instance followed by the average energy $E_{avg}$, where in the classical methods $E_{avg}$ is computed by taking the average outcome over 100 samples of the same instance and for the quantum  by computing the expectation value of the energy of the final quantum state. Motivated by \cref{def:MCOP}, we choose \(\alpha \) to guarantee that our solution is close to the optimal solution of the instances, in particular with \(\alpha = 0.9\) it will give solutions between the ground state energy and the first excited state energy for the SKOP instances considered (see  \cref{fig:ratio_plot}). However, if we were to include larger system sizes while aiming to keep the solutions within this range, the value of \(\alpha\) would need to increase. 

Now, for optimal solutions generated by the Grover Adaptive Search (GAS), the number of steps $T$ are the number of oracle calls needed to attain a success probability of \(1 - \varepsilon\), with \(\varepsilon = 0.016\). 

A key aspect of our benchmarking strategy is predicting the asymptotic scaling, represented by the $c$ values in \cref{fig:main_plot} (where $N=2^n$ denotes the number of spin configurations for systems with $n$ spins), of the steps needed to reach either the optimal solution or the 0.9-AEAR. These $c$ values are obtained through curve fitting (by scipy curve fitting function) from our numerical analysis shown in \cref{fig:main_plot}.

Brute force (BF) is crucial in our benchmark as it is the only method with provable scaling, and our numerical results in \cref{fig:main_plot} align with expectations, showing linear scaling. Similar behaviour is anticipated for random sampling; however, for the 0.9-AEAR in the SK model, the number of approximate solutions significantly exceeds the optimal solutions. Additionally, while the GAS shows a quadratic improvement over BF—consistent with Grover’s algorithm’s expected speedup—it is a nondeterministic method based on the Grover oracle, making the quadratic speedup over BF less predictable when the number of solutions is unknown.

Despite deriving results for large system sizes from small system data, the key characteristics of the optimization problem remain unchanged, as they are defined by the structure of the SK Hamiltonian. 

Finally, the numerical scaling of each classical and quantum algorithm provides an estimate of the average number of steps needed to find either the approximate or optimal solution to the cost function, independent of hardware. Again we should stress that the number of steps prediction for the $\alpha=0.9$ is a prediction for solutions between the ground and first state excited state energy of the SK Hamiltonian, which will corresponded to a higher $\alpha$ value for larger systems sizes. The benchmark is completed by estimating the clock time, reported in \cref{tab:sk_estimates}, for each heuristic, revealing a clear distinction between classical and quantum hardware.

\begin{table*}[t]
\begin{tabular}{ c|c|c|c}
\textbf{ Classical and Quantum}  & \textbf{Spin size} & \textbf{Days}  &   \textbf{Galactic} \\
  \textbf{methods} & $\mathbf{n}$ &  &\textbf{year}  \\
\hline
 Brute force - Classical unstructured& 64&  $1.67\!\times\! 10^7$  & $1.99 \times 10^{-4}$\\ 
 (Opt. solution)& 128 & $6.08\!\times\! 10^{26}$ & $7.24 \times 10^{15}$  \\
 \hline
  Random sampling - Classical unstructured& 64 & $1.17\!\times\! 10^3$ & $-$ \\ 
  (Approx. solution)& 128 & $5.00\!\times\! 10^{16}$ & $5.95 \times 10^{5}$   \\    
\hline Metropolis-Hastings - Classical structured  & 64 &  $6.91 \times 10^{-7}$ & $-$\\ 
(Approx. solution) & 128  &$1.70\times 10^{-3}$ & $-$  \\ 
\hline 
 GAS - Quantum unstructured & 64 & $2.00\!\times\! 10^9$ & $2.38 \times 10^{-2}$ \\ 
 (Opt. solution) & 128  & $1.42\!\times\! 10^{20}$ &$1.69 \times 10^{9}$  \\    
\hline LHPST  - Quantum structured & 64 & $2.96\times 10^1$& $-$   \\ 
 (Approx. solution)& 128  & $4.88\!\times\! 10^{5}$ & $5.81 \times 10^{-6}$\\ 
\hline 
\end{tabular}
\caption[Resource estimates for the SK problem]{\label{tab:sk_estimates}
\textit{Estimates of time.} We provide the total time required by all methods for spin sizes of $64$ and $128$, with estimates expressed in both days and Galactic years (1 Galactic year = $2.3 \times 10^8$ years). For the classical heuristics, the total time is estimated by the total floating-point operations needed to achieve either the optimal or approximate solutions of the SK Hamiltonian, see \cref{tab:Float_op}. This estimation assumes the computational power of an iPhone 12 ($11$ TFLOPS). For the quantum methods, the estimated time is based on the Toffoli count, as shown in \cref{tab:T_count}. These tables are available in \cref{app:comp_analy}.  The estimated number of all error-correction overheads are computed assuming a single Toffoli factory using state distillation as outlined in \cite{Gidney2019} and reported in \cite{sanders2020compilation}. The target success probability is set to $0.9$.}
\end{table*}

Our main results are given in \Cref{fig:main_plot},
which summarizes the simulated numerical performance of selected classical and quantum heuristics
on small random instances of the SK model.
Our full methodology and detailed results are given in \cref{sec:optimization_algorithms}.
We take the fitted value of $c$ as an indication of the asymptotic scaling of the average
case performance of the heuristic, noting that we expect all approaches to scale exponentially.
We observe, as expected, that the scaling of the Brute Force approach is roughly $2^n$ ($c\approx 1$)
and the scaling of Random Sampling is roughly $2^{n/\sqrt{2}}$ ($c\approx 1/\sqrt{2}$).

By contrast, Grover Adaptive Search scales as $2^{n/2}$ ($c \approx 0.5$).
Whereas these three results clearly indicate an advantage for the quantum heuristic over classical competitors,
we emphasize that all three are \emph{unstructured} approaches.

The comparison between our chosen classical and quantum \emph{structured} approaches tell a different story:
the fitted $c$ values are quite similar, with a small advantage being given to the classical version of Metropolis-Hastings.
We hasten to emphasize that our approach is not able to determine that classical is actually better than quantum,
only that our numerical analysis does not provide clear evidence one way or another.

When evaluating the potential of classical and future quantum hardware capabilities, we refer back to Ref.~\cite{sanders2020compilation} to estimate the time quantum devices would take to solve the SK problem for two different spin sizes, compared to classical methods executed on classical hardware. One key issue is the substantial use of Toffoli gates, which makes the GAS method impractical for large-scale optimization problems. For example, solving the SK Hamiltonian with $n = 64$ spins could require computational times on the order of a galactic year.

In contrast, the LHPST method shows a much more efficient solution for space exploration, providing greater-than-quadratic speedup over GAS and reducing the computational time for less than one month rather than years. However, when comparing LHPST to classical simulated annealing (MH), assuming that we have only access to the computational power of a smartphone (measured in TFLOPS) rather than the best supercomputers (measured in exaFLOPS), the MH algorithm can solve the same problem in seconds, while LHPST may still take weeks.

We note that in this work we do not consider the QAOA due to the cost of optimizing the angles of the ansatz, for each instance and dimension considered. The performance of QAOA for optimization of the SK model and as a search algorithm has been studied in Refs.~\cite{Farhi2022quantumapproximateSK, JiangRieffelWang2017,Morales2018qaoagrover,Bartschi2020qaoagrover}. Another important quantum technique not included here is the quantum simulated annealing \cite{kirkpatrick1983optimization}, which includes the spectral amplification method \cite{boixo2015fast} that gives quadratic quantum speedup for simulated annealing due to its spectral gap dependence. Since in this approach, for each instance of the SK, we have to construct the Hamiltonian associated with the classical transition matrix for the classical Markov chain, the gap will be instance-dependent. Therefore, given that here we consider hundreds of different instances, such analysis would be infeasible.

In summary, our benchmarking approach establishes a systematic framework for evaluating the efficiency of classical and quantum optimization methods in solving the spin glass problem. By focusing on the time required to reach a specified solution quality, we enable a more detailed comparison that accounts for the practical limitations of quantum algorithms. Our results emphasize that while quantum algorithms like LHPST demonstrate significant potential, their performance must be assessed in the context of the available computational resources and the specific problem characteristics.

Interestingly, our results did not show the quadratic speedup of the quantum heuristic over its classical counterpart, as expected by \cite{sanders2020compilation}, based on results from \cite{boixo2015fast}. The LHPST algorithm offers an alternative way to access the same gap dependence described in \cite{boixo2015fast}, which indicates a quadratic speedup relative to the classical Metropolis-Hastings (MH) algorithm when both start at the same temperature, share the same final target temperature, and employ their respective classical and quantum walk operators with the fixed target temperature. However, since we are using different temperatures at each walk step during the optimization  which improves the convergence behaviour, we are not waiting for the termalization to the Gibbs state for the intermediate temperatures to happen, so the quadratic speedup is not guaranteed, (for further details see \cref{sec:clas-MH} for the MH and \cref{sec:lhpst} for the LHPST). 

Future work could explore different values for  $\alpha$ in the $\alpha$-AEAR; we might find that classical or quantum-structured methods offer better scaling behaviour when both are contrasted. Additionally, we could explore a broader range of cost functions and heuristics to deepen our understanding of optimization algorithms. Possible cost functions include Ising models with varying interaction terms, random optimization landscapes, and multi-modal functions, which could yield further insights into the performance of both classical and quantum methods. 

\emph{Acknowledgements:--} We thank Stephen Jordan for his helpful comments. 
MESM acknowledges support from the U.S. Department of Defense through a QuICS Hartree Fellowship.
YRS acknowledges support from the New South Wales Defence Innovation Network 
as well as the Defense Advanced Research Projects Agency under Contract No.~HR001122C0074.
Any opinions, findings and conclusions or recommendations expressed in this material are those of the authors and do not necessarily reflect the views of the Defense Advanced Research Projects Agency.


%

\appendix 

\section{Classical to Quantum Mapping of the SK Hamiltonian}
\label{app:classictoQuan}

Our cost function is represented by the following classical Hamiltonian:
\begin{equation}
\label{eq:SKs}
E(\mathbf{s}) = \sum_{i \neq j}^{n} w_{ij} s_i s_j,
\end{equation}
where $w_{ij} = w_{ji} = \pm 1$ are the random coupling constants that encode the problem instance, and $n$ represents the total number of spins.

In the context of quantum heuristic methods, the corresponding version of the SK Hamiltonian is given by:
\begin{equation}
H_E = \sum_{i \neq j} w_{ij} Z_i Z_j.
\label{eq:QSk}
\end{equation}
This expression defines a diagonal Hamiltonian in the computational basis \(\ket{\mathbf{x}} = \ket{x_1} \cdots \ket{x_n} \in \mathcal{H}^{2^n}\), satisfying the relation $H_E \ket{\mathbf{x}} = E(\mathbf{x}) \ket{\mathbf{x}}$, where $E(\mathbf{x})$ is provided by \cref{eq:SKs}.

\section{Optimization methods}\label{sec:optimization_algorithms}
\subsection{Classical methods}
\subsubsection{Brute Force and Random Sampling}
The simplest methods that do not involve any structure in the search applied to find the exact and the $\alpha$-average approximate solutions of the SK model are the brute force (BF) and random sampling (RS), respectively.

In the BF approach, we check the candidate solutions one by one, always following the same sequence, i.e., there is neither randomness in the starting point nor the sequence used.  We tested the BF performance by counting how many iterations the algorithms must be done until we get the ground state solution, which is previously computed.

In RS, we only have to randomly generate values $x$ from a uniform distribution, i.e., $x=\left\lfloor{U[1,2^n]}\right \rfloor$, where $n$ is the total number of spins used in our Hamiltonian \Cref{eq:SKs}, with a subsequent calculation of $E(x)$. We do it $T-1$ times, from a randomly chosen starting point, $x_{start}$, where at each time we always compare the new solution with the previous one and then keep with the value $x$ that has the minimum energy value.

\subsubsection{Metropolis-Hastings: a randomized method with structured search}\label{sec:clas-MH}

The simulated annealing, also known as the Metropolis-Hastings algorithm, systematically explores the solution space by generating candidate solutions and accepting or rejecting them based on a probabilistic criterion (related to the cost function). This is not a brute-force search but rather a guided exploration driven by the structure of the problem through the proposal distribution. For this reason, we say that it is a structure-based search. 
Regardless of the cost function, can be summarized by the following sequence of steps.

\begin{algorithm}[H]
  \caption{Simulated Annealing}
   \label{Alg:classic_Met}
   \begin{algorithmic}[1]
   \State \textbf{Input}: $(x_{start}=\left\lfloor{U[1,2^n]}\right \rfloor,\beta_{T}, T)$\quad  $x\leftarrow x_{start}$, \; $\beta_{1} \leftarrow 0$
    \For{$i=1,2,\cdots,T$} 
      \State  Generate ($y$ from $x$)
      \State  \textbf{if} $f(y)\leq f(x)$ \textbf{do}
      \State  \quad $x\leftarrow y$
      \State  \textbf{else}  \textbf{do}
      \State  \quad \textbf{if} $\exp\{\beta_i\left(f(x)-f(y)\right)\} >[0,1)$ \textbf{do}
      \State  \quad \quad $x\leftarrow y$
      \State $\beta_{i+1}\leftarrow g(\beta_i)$
    \EndFor
    \end{algorithmic}
\end{algorithm}

In the algorithm above, $T$ represents the total number of steps executed by the heuristic algorithm. In the simulated annealing process, the parameter $\beta_i$ functions as the inverse temperature at step i. The variable $y \in \mathcal{N}(x)$ indicates that $y$ is a neighbour of $x$, with all neighbours having the same size, defined by $|\mathcal{N}(x)| = M$. Furthermore, the function $f: \{0,1\}^n \rightarrow \mathbb{R}$ denotes the cost function for the given optimization problem based on the bit string representation of $x$. In our case, this cost function corresponds to the Sherrington-Kirkpatrick Hamiltonian such that $f(x) = E(x)$. There are several strategies for generating $y$ from $x$ and updating $\beta$, given by the function $g(\cdot)$, and we discuss our specific choices in \Cref{app:neigh_sched}.

The algorithm above treats the Ising model as a Markov chain, which in stochastics is a chain of events whose probabilities depend only on the spin configurations prior to that event. The probabilities $p_x$ of the system to be in the certain configuration $x$ and the probabilities $\mathcal{W}_{yx}$ for the transition from one configuration $x$. 

In our algorithm, we have $\mathcal{W}_{xy}$, \textit{transition probability}, represented from steps $4$ to $8$ in \Cref{Alg:classic_Met} for the SK Hamiltonian which can be restated as 
\begin{equation}
\label{eq:transi_probSK}
\mathcal{W}^{(\beta_i)}_{yx} = \left\{ \begin{array} {ll}
T_{yx} &\quad \text{if } y\in \mathcal{N}(x) \\
1-\sum_{z\neq x}T_{zx} &\quad\text{if } y = x\\
0 &\quad\text{otherwise, }
\end{array}\right.
\end{equation}
where 
\begin{equation}
\label{eq:accep_Prob}
T_{yx} = \frac{1}{M}\min{\left(1,e^{\beta_i(E(x)-E(y))}\right)},    
\end{equation}
for $x\neq y$, is the \textit{acceptance probability matrix}.

The main idea of having $\mathcal{W}^{(\beta_i)}$, also known as the \textit{random walk} matrix \Cref{eq:transi_probSK}, is that by sampling from a distribution $p$ to the distribution $p'=\mathcal{W}^{(\beta_i)}p$, where its elements are given by $p'_y=\sum_x\mathcal{W}_{xy}^{(\beta_i)}p_x$, we will achieve the equilibrium distribution, $\pi^{\beta_{i}} =\mathcal{W}\pi^{\beta_{i}}$  which is characterized by the Boltzmann distribution, also called by \textit{Gibbs distribution}
\begin{equation}
\label{eq:class_Gibs}
\pi^{(\beta_{i})}_x=\frac{\exp\{-\beta_iE(x)\}}{\sum_y\exp\{-\beta_i E(y)\}},
\end{equation}
where the detailed balance $\mathcal{W}_{yx}\pi_x^{(\beta_i)} = \mathcal{W}_{xy}\pi_y^{(\beta_i)}$, is satisfied. 
 
The way that we quantify how fast the equilibrium is achieved in the Markov chain method is by looking at its \textit{mixing time}, which is governed by the inverse spectral gap $0\leq \Delta\leq 1$ of $\mathcal{W}$, the difference between its two largest eigenvalues of the random walk operator. Therefore, we can associate a sequence of random walk operator $\{\mathcal{W}^{(\beta_{1})}, \cdots,  \mathcal{W}^{(\beta_{T})} \}$ for each inverse temperature, where $\beta_1 < \cdots < \beta_T$. So, the estimated runtime of the MH algorithm for getting the Gibbs distribution at $\beta_T$ is the product of the mixing time and the time required to implement a single step of the walk,
\begin{equation}
\label{eq:MH_cost}
 C \sum_{i=1}^T\frac{1}{\Delta(\beta_i)},   
\end{equation}
where $C$ is a constant that considers the time of the walk step, which is dimensional dependent, and $\Delta(\beta_i)$ is the gap of the walk with the inverse temperature given by $\beta_i$.

Since we do not analyze the gaps of $\mathcal{W}^{(\beta_{i})}$ due to the numerous instances of the SK problem considered and because we are most interested in the $\alpha$-AEAR, we always update the new value of $\beta_i$, see \cref{app:neigh_sched}, after a single step of the walk. So, the total cost will be just $CT$ (without convergence guarantee to the Gibbs distribution at the final inverse temperature $\beta_T$).

\subsection{Quantum methods}
\subsubsection{Grover Adaptive Search}
\label{sec:grover_search}

Likewise, in the Grover problem that has a unique marked state, in the case of $t$ solutions, we can also split the equal superposition state with $n$ qubits $\ket{\psi_0} = 1/\sqrt{2^n}\sum_i\ket{x_i}$ in two subspaces. One with solutions, i.e., $x_i \in S$ and the other that does not satisfy the solution, $x_i \notin S$. 
 
When the Grover method, without knowing the number of solutions, has to be applied in practice, it becomes an interactive method, also known as Grover adaptive search, which will be used for our optimization problem. 

We evaluate the performance of GAS by counting the accumulated steps until we get $1-\varepsilon$ as the probability of success in finding one of the possible solutions, where $\varepsilon$ is the tolerance error, which we considered $\varepsilon=0.016$. 
One way of implementing GAS is using \Cref{Alg:Grover_search}. To perform the numerical tests, we have simplified the algorithm by first finding the solution to each instance and using it to implement the standard Grover algorithm. Therefore, the numbers in \Cref{fig:main_plot} for GAS should be considered as an optimistic estimation of the total number of steps. 

Although there exist improved results of the Grover adaptive method, we decided to adopt the original proposal from Ref.~\cite{tightbounds}, since the improvement results in Ref.~\cite{liu2010using}, will not change the overall scaling of the Grover method, but only will give a smaller constant factor. Notwithstanding, for practical purposes, we consider the proposal from Refs.~\cite{baritompa2005grover,gilliam2021grover}, which is explained later here.

In \Cref{Alg:Grover_search}, $T$ is used as the accumulator variable. The idea here is that every time that you apply the oracle $k$ times, which is selected randomly from a uniform distribution from 1 to $m$, represented by $\left\lfloor{U[1,m]}\right \rfloor$, and the algorithm dooes not succeed we add the number $k$ in the accumulator variable. Notice that each instance of the problem requires a new oracle $G$. 

\begin{algorithm}[H]
  \caption{Grover Adaptive Search}
  \label{Alg:Grover_search}
   \begin{algorithmic}[1]
   \State \textbf{Input}: $(\lambda,n,m,k,\varepsilon,T,G)$\quad $\lambda=6/5$, \; $m\leftarrow \lambda$, \; $k\leftarrow 1$, \: $T\leftarrow 0$
   \State $\ket{\psi_0}=H^{\otimes n}\ket{0^n}$
   \State $G^k\ket{\psi_0}$
   \State $T\leftarrow T +k$
    \While {$\sum_{i\in S}|\bra{i}G^k\ket{\psi_0}|^2 \leq 1-\varepsilon$} 
    \State $\ket{\psi_0}=H^{\otimes n}\ket{0^n}$
    \State  $m\leftarrow min(\lambda m, \sqrt{2^n})$
    \State  $k\leftarrow \left \lfloor{U[1,m]}\right \rfloor$
    \State  $G^k\ket{\psi_0}$
    \State $T\leftarrow T +k$
    \EndWhile
    \end{algorithmic}
\end{algorithm}

\paragraph{Circuit encoding for GAS.}

Given an objective function such as in \cref{eq:QSk}, the idea is to generate a superposition of all possible input values $\x\in\{0,1\}^n$, evaluate the function at each of these inputs and then mark the inputs whose value are smaller than some fixed integer $y$. One can use Grover for this search and then repeat the search with the newly obtained value. In this section, we provide some details on this method, but for a more in-depth discussion, see \cite{baritompa2005grover,gilliam2021grover}, where in particular in Ref.~\cite{gilliam2021grover}, the authors consider using GAS on quadratic unconstrained binary problems.

As in Ref.~\cite{gilliam2021grover}, we consider two registers $\ket{\x}_{\rin}\ket{\z}_{\rval}$ where $\x\in \{0,1\}^n$ and $\z\in\{0,1\}^\ell$. The register $\rin$ stores the input to the objective function $E(\x)$ and the register $\rval$ stores the value of the objective function $E(\x)$. 

The encoding of an integer follows the construction in Figure 3 of \cite{gilliam2021grover}. This encoding is based on the quantum Fourier transform. The integer $y$ is encoded in a bitstring $\y\in\{0,1\}^\ell$ where the leftmost bit encodes the sign of the number.

\paragraph{Implementing operator $A$ and oracle $O$.}
The GAS algorithm is based on an oracle $A$ which prepares an initial superposition and an oracle $O$ which checks whether the energy is lowered. The construction of oracle $A$ and $O$ follows \cite{gilliam2021grover}.

The oracle $O$ is simple to implement in this encoding as it only needs to check the leftmost qubit, which indicates the sign of $E(\x)-y$ and multiply by $-1$ accordingly. The Grover iteration dependent on value $y$ for GAS is given by $\tilde{G}_y=A_y D A_y^\dagger O $, where $D=I-2\ketbra{0^n}{0^n}$ .

Now, we give a gate count for implementing the Grover iteration $\tilde{G}_y$. In the framework we have given, the oracle $O$ can be implemented by a $Z$ gate on the corresponding qubit, indicating the sign of the energy difference. To count the Toffoli gates from $D$, we note that the only contribution is from the CNOT with $n$ control qubits, which can be implemented with $2(n-1)$ Toffoli gates and a single CNOT. Due to the encoding used, the oracle does not require Toffoli gates. For the operator $A_y$, we use the result from \cite{PRXQuantum.1.020312}, which gives a Toffoli count of $n^2$ (see the direct energy oracle in Table I in \cite{PRXQuantum.1.020312}). This does not include the offset of $y$ in the energy, which will only contribute to a constant factor in the Toffoli count. The Toffoli count is summarized in \Cref{tab:gate_count}.

\begin{table}
\centering
\begin{center}
\begin{tabular}{ |c|c| } 
 \hline
\textbf{Operator} & \textbf{Toffoli count} \\ 
 \hline
$O$ & 0
\\ 
$D$ & $2(n-1)$
\\ 
$A_y$ & $n^2$
\\
\hline
\end{tabular}
\end{center}
\caption{\textit{Number of Toffoli gates for each operator in the Grover iteration}. Operator $O$ refers to the oracle used in the Grover iteration $\tilde{G}_y$ implemented as a $Z$ gate on the corresponding qubit. Operator $D$ refers to the diffusion operator, also known as the inversion about the mean. Operator $A_y$ acts as a state preparation operator.}
\label{tab:gate_count}
\end{table}

\subsubsection{Quantum Metropolis-Hastings: LHPST-qubitized walk-based quantum simulated annealing}\label{sec:lhpst}
Here, we explored the quantum Metropolis-Hastings algorithm proposed first by Lemieux, Heim, Poulin, Svore, and Troyerin in Ref.~\cite{lemieux2020efficient} and improved in Ref.~\cite{sanders2020compilation}. The technique given in Ref.~\cite{lemieux2020efficient} is a special case of the Szegedy quantum walk \cite{szegedy2004quantum}, the quantum counterpart of the random walk. While in the classical method, we have at the end the equilibrium distribution given by the $\Cref{eq:class_Gibs}$ in its quantum counterpart, the idea is to construct the quantum walk, $U_{\mathcal{W}}$, a unitary operator, such that it produces the coherent version of the Gibbs state, $\ket{\pi} = \sum_x\sqrt{\pi^{(\beta_i)}(x)}\ket{x}$ also known as the quantum Gibbs state.

In Ref.~\cite{lemieux2020efficient}, authors propose a less expensive circuit implementation of the quantum walk that requires three registers $\ket{x}_S\ket{z}_M\ket{b}_C$, namely the \textit{System}, \textit{Move}, and \textit{Coin} registers, and composed by four unitary operators,
 \begin{equation}
 \label{eq:walk_lhpst}
 \tilde{U}_{\mathcal{W}} = RV^{\dagger}B^{\dagger}FBV,    
 \end{equation}
 where is the operator $B$, which depends on the inverse temperature and $F$ on the neighbouring structure; where the details about the operators used can be find in \cite{lemieux2020efficient,sanders2020compilation}. 

Analogous to the classical annealing, we can associate a sequence of quantum random walk operator $\{\tilde{U}_{\mathcal{W}^{(\beta_{1})}}, \cdots,  \tilde{U}_{\mathcal{W}^{(\beta_{T})}} \}$ for each inverse temperature, where $\beta_1 < \cdots < \beta_T$. Like in the classical case, the number of steps to achieve the quantum Gibbs state depends on the inverse gap $\delta$ of the quantum walk, where the gap is for the eigenvalues of the unitary operator $U_{\mathcal{W}}$. However, given a random walk $\mathcal{W}$ that has a gap  $\Delta$, its quantum walk has a gap $\delta \geq \sqrt{\Delta}$, so we will have at least a quadratic faster convergence for the quantum Gibbs distribution in opposed the Gibbs distribution given by the classical walk, for each $\beta$, where the overall complexity is given by
\begin{equation}
\mathcal{C}\sum_{j=1}^T \frac{1}{\delta(\beta_j)}, \end{equation}
where $\delta(\beta_j)$ is the gap of $\tilde{U}_{\mathcal{W}^{(\beta_{1})}}$ and $\mathcal{C}$ and is the time required to perform a single quantum walk step in the quantum computer.

As with the classical random walk, we do not analyse the gaps of $\tilde{U}_{\mathcal{W}^{(\beta_{i})}}$ due to the numerous instances of the SK problem considered and because we are also only interested in the $\alpha$-average approximation of the ground state. Thus, we always update the new value of $\beta_i$ after a single step of the walk, \Cref{Alg:Quantum_Met}. So here, the total cost will be $\mathcal{C}T$. We also say that the LHPST gives a structure search in opposition to the GAS method for the same arguments that we give for the MH algorithm, and we used the same type of strategy to move from one candidate solution and how we update the inverse temperature at each step, \Cref{app:neigh_sched}.

\begin{algorithm}[H]
  \caption{LHPST}
   \label{Alg:Quantum_Met}
   \begin{algorithmic}[1]
   \State \textbf{Input}: $(R,V,F,\beta_{T}, T)$\quad  $\beta_{1} \leftarrow 0$
   \State $\ket{\psi_0}=\ket{0}_S\ket{0}_M\ket{0}_C$
    \For{$i=1,2,\cdots,T$} 
        \State Compute $B^{(\beta_i)}$
        \State $\tilde{U}_{{\mathcal{W}^{(\beta_i)}}} = RV^{\dagger}B^{(\beta_i)\dagger}FB^{(\beta_i)}V$
        \State $\ket{\psi_i} =\tilde{U}_{{\mathcal{W}^{(\beta_i)}}} \ket{\psi_{i-1}}$
      \State $\beta_{i+1}\leftarrow g(\beta_{i})$
    \EndFor
    \end{algorithmic}
\end{algorithm}

\section{The neighbouring and scheduling schemes} \label{app:neigh_sched}
 
In this work, $y$ differs from $x$ by exactly one bit-flip in all cases. In other words, the value $y$ is computed from the function $g^{(j)}: \{0,1\}^n \rightarrow \{0,1\}^n$ that flips the bit component $x_j$ of $x$, $x[x_j]$, to $\bar{x}_j$, i.e., we have 
\begin{equation}
\label{eq:bit_flipF}
g_j: x[x_j] \rightarrow x[\bar{x}_j] = y,
\end{equation}
where $\bar{x}_j = x_j \oplus 1$ and the bit $j=0,\cdots,n-1$ is selected with probability $1/n$. That is, we have our random walk on a $n$-regular graph $G$, $|\mathcal{N}(x)|=n, \,  \forall x \in G$. The one-bit-flip difference is also considered in the LHPST algorithm, where more details on how the quantum algorithm works are given in \cref{sec:lhpst} .

Although the one bit-flip difference allows us to move the values $x$ for a large amount, for example, by flipping the most significant bit of $x$ that has a binary representation, being trapped into local optimal values of the cost function is likely to happen at larger values of $\beta$. So here, for the classical MH, we also added a feature that $x$ could be tunnelled, with a low probability, from local traps by randomly generating $y$ without any constraints, but for the analysis purpose of our MH, we do not include such a possible move. 

Now, for the updating rule of the beta parameter, we apply linear scheduling as follows
\begin{equation}
g(\beta_i) = \beta_{i} + \beta_T/T.    
\end{equation}
Since the cost functions are randomly generated and we do not have previous information about the gap value of \Cref{eq:SKs}, our final value of the inverse temperature is always the same $\beta_T=1$, and the question is if such an approach will give us a good convergence.

\section{Complexity Analysis: Classical and Quantum Operation Counts}
\label{app:comp_analy}

\subsection{Computational cost on quantum hardware}

To estimate the time of a single step of the quantum methods, we consider the computational cost of superconducting qubit surface code processors without significant improvements in implementing the surface code, as it is considered in \cite{sanders2020compilation}. This is then translated to a Toffoli count, the costly gate for the fault-tolerant quantum computer considered in this work. 

For the quantum heuristics considered in this paper, the total number of Toffoli gates for a single step as a function of a spin system size is already given in \cite{sanders2020compilation} so the total Toffoli count to solve either the optimal or the approximated solution is then completed by using our interpolation results from \Cref{fig:main_plot} that gives the total number of steps as a function of a spin system size, which is quantified in \Cref{tab:T_count}. For simplicity, we only consider the time for the total number of steps, so we do not consider the order of repetitions or the measurement time to extract the solution, which will give extra overheads in the computational time. 

\begin{table}[t]
\centering
\begin{center}
\begin{tabular}{ |c|c|} 
 \hline
\textbf{Quantum method}  &  \textbf{Total steps }    \\
& $\mathbf{\times}$ \textbf{T count per steps}\\
 \hline
GAS & $4.27\times 2^{0.50n}$ \\
(Opt. solution)& $\times(2n^2+n+ \mathcal{O}(\log n))$
\\
\hline
LHPST & $8.88\times 2^{0.21n}$ \\
(Approx. solution)& $\times(5n+11\log n +98)$\\
\hline
\end{tabular}
\end{center}
\caption{\textit{Toffoli count for GAS and LHPST.} The Toffoli count is computed for the optimal solution in the case of Grover Adaptive Search (GAS) and for the approximate solution in the case of the Quantum Metropolis-Hastings algorithm (LHPST), as a function of the system size for the spin glass Hamiltonian.}
\label{tab:T_count}
\end{table}

\subsection{Computational cost on classical hardware}

\begin{figure}
    \centering
    \includegraphics[width=0.35\textwidth]{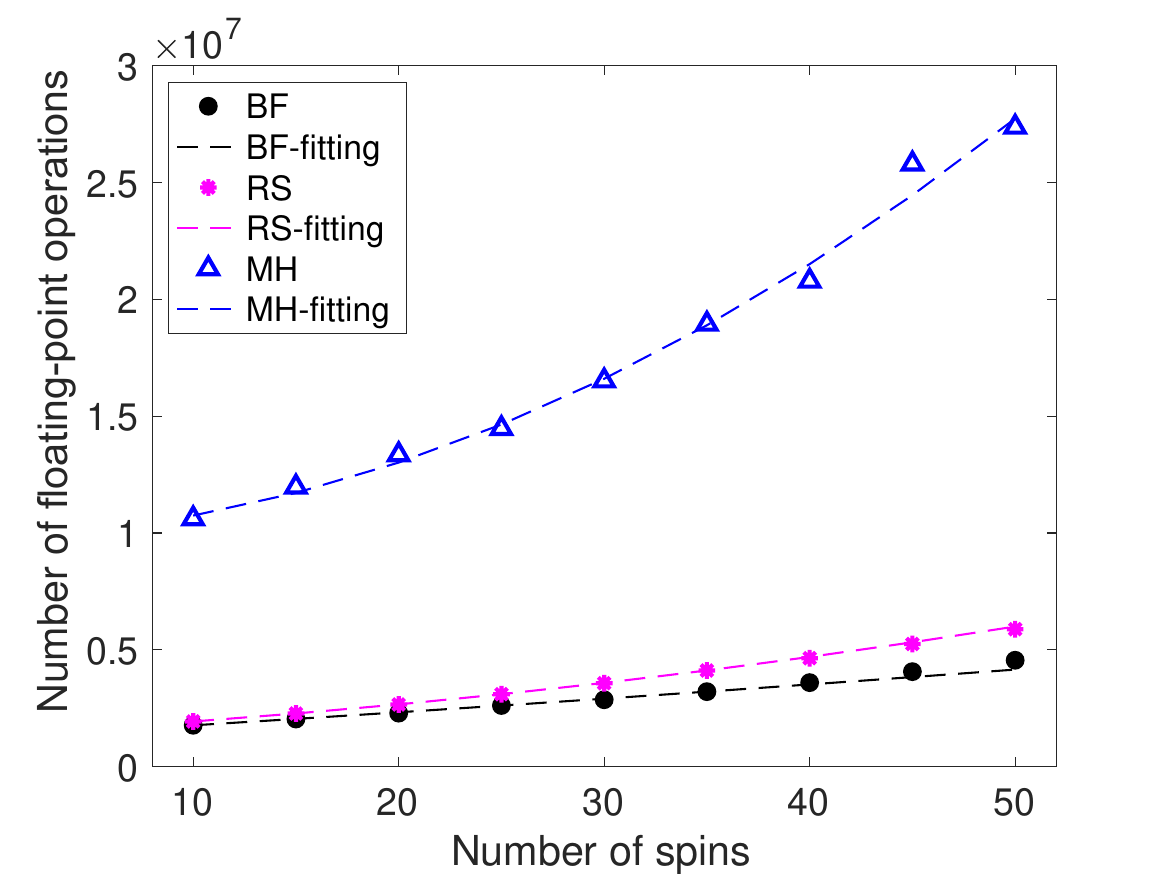}\\
    \caption{\textit{Number of floating-point operations and fitting results for each classical method}. The estimated FLOPs are provided for spin system sizes ranging from $10$ to $50$. These values represent the arithmetic mean of the floating-point operations across all instances of the SK Hamiltonian for each system size. The analysis highlights the computational complexity associated with each classical approach.}
    \label{fig:FOPs}
\end{figure}

In classical computation, we first consider the number of floating-point operations (FLOPs) for each call of the main subroutine explored to estimate the total time at each step. Since all algorithms used, namely BF, RS, and MH, only have basic operations involved: multiplication, addition, and subtraction, we expected the following worst-case asymptotically scaling $\mathcal{O}(n^2)$ for the number of FLOPs as the system size $n$. We say worst-case scaling because, despite the usual expectation of $\mathcal{O}(n^2)$ operations in the multiplication between two $n$-bit integers, the best-known results for multiplication, Ref.~\cite{harvey2021integer} has $\mathcal{O}(n\log(n))$ complexity scaling.  As we can expect, we can see in \Cref{fig:FOPs} that the number of FLOPs increases with the algorithm's complexity; the BF is the simplest method and involves less computation for each step.

Once we have the number of FLOPs as a function of the number of spins, inferred from \Cref{fig:FOPs}, we need to include the interpolation results for the total number of steps from \Cref{fig:main_plot}, which is given in \cref{tab:Float_op} and \Cref{tab:sk_estimates}. We then estimated the total time by considering the computational power given by the iPhone 12, which, accordingly with Apple \cite{apple}, is $11$ Tera ($10^{12}$) floating-point operations per second (TFLOPS). As with quantum methods, we do not account for the number of repetitions needed in the randomized heuristic methods to achieve an approximate solution, which may introduce overhead. However, based on our numerical results, we estimate that no more than two repetitions are required to reach the predicted outcomes.

\begin{table}[h]
\centering
\begin{center}
\begin{tabular}{ |c|c|} 
 \hline
\textbf{Classical method}  &  \textbf{Total steps}    \\
  &  $\mathbf{\times}$ \textbf{floating-point operations per steps}  \\
 \hline
 Brute force & $0.34\times2^{0.94n}$ 
\\
 (Opt. solution)& $\times(1.3\times 10^2  n^2+5.2\times 10^4n+ 1.2\times10^6)$ \\
\hline
Random sampling & $3.05\times 2^{0.61n}$ 
\\
 (Approx. solution)& $\times\,(9.5 \times 10^2n^2+4.5\times 10^5n+ 1.4\times10^6)$\\
\hline
Metropolis-Hastings & $14.09\times 2^{0.17n}$ \\
(Approx. solution)& $\times\,(6.6 \times 10^3n^2 + 2.9\times 10^4n+ 9.8\times10^6)$ \\
\hline
\end{tabular}
\end{center}
\caption{\textit{Floating-point operations in the classical algorithms.} The total number of floating-point operations is estimated for classical heuristic algorithms as a function of the total number of spins in the Sherrington-Kirkpatrick model. This estimation provides insight into the computational complexity and efficiency of these algorithms across varying spin sizes, highlighting the scalability of classical methods in solving complex optimization problems.}
\label{tab:Float_op}
\end{table}

\section{Numerical analysis: supporting material}\label{app:support}
\begin{figure}
\centering
\includegraphics[width = 0.45\textwidth]{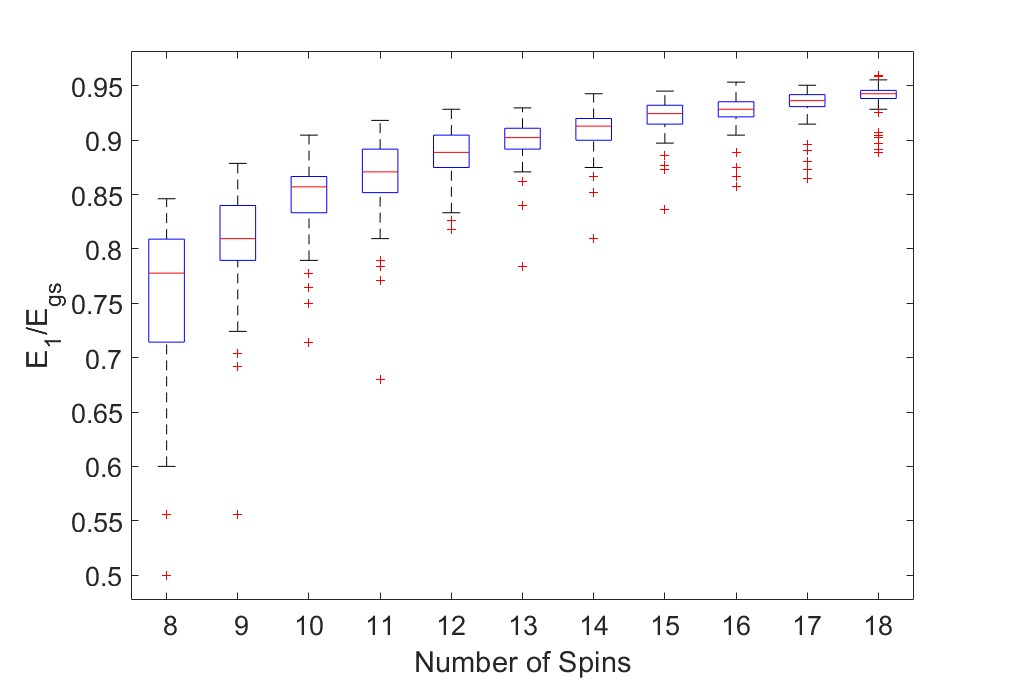}\\
    \caption{\textit{Box plot results of the ratio gap, $E_1/E_{gs}$,  over the $100$ randomly generated SK instances considered in the numerical analysis, from 8 to 13 spins, including the spin sizes until 18.}
    }
    \label{fig:ratio_plot}
\end{figure}

\begin{figure}
    \centering
    \begin{tabular}{@{}p{0.7\linewidth}}
    \subfigimg[width=\linewidth]{(a)}{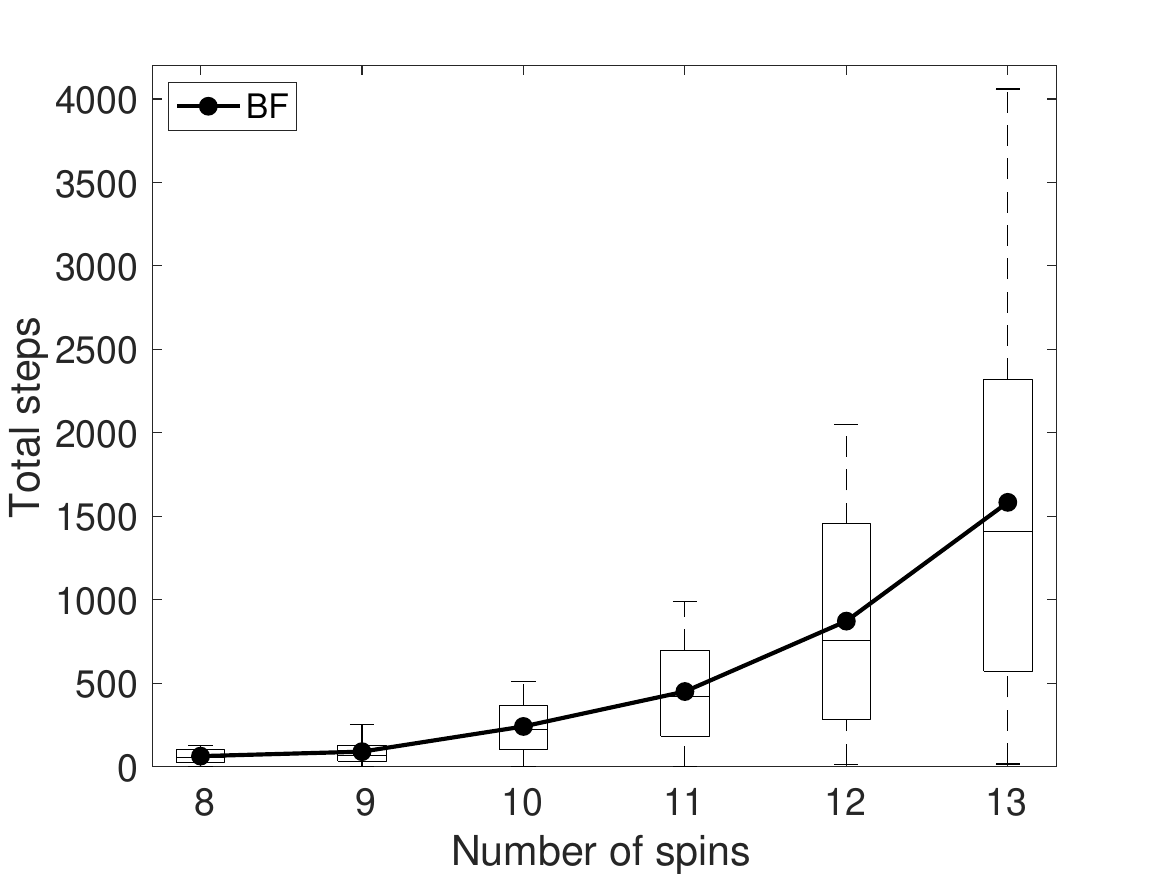} \\
    \subfigimg[width=\linewidth]{(b)}{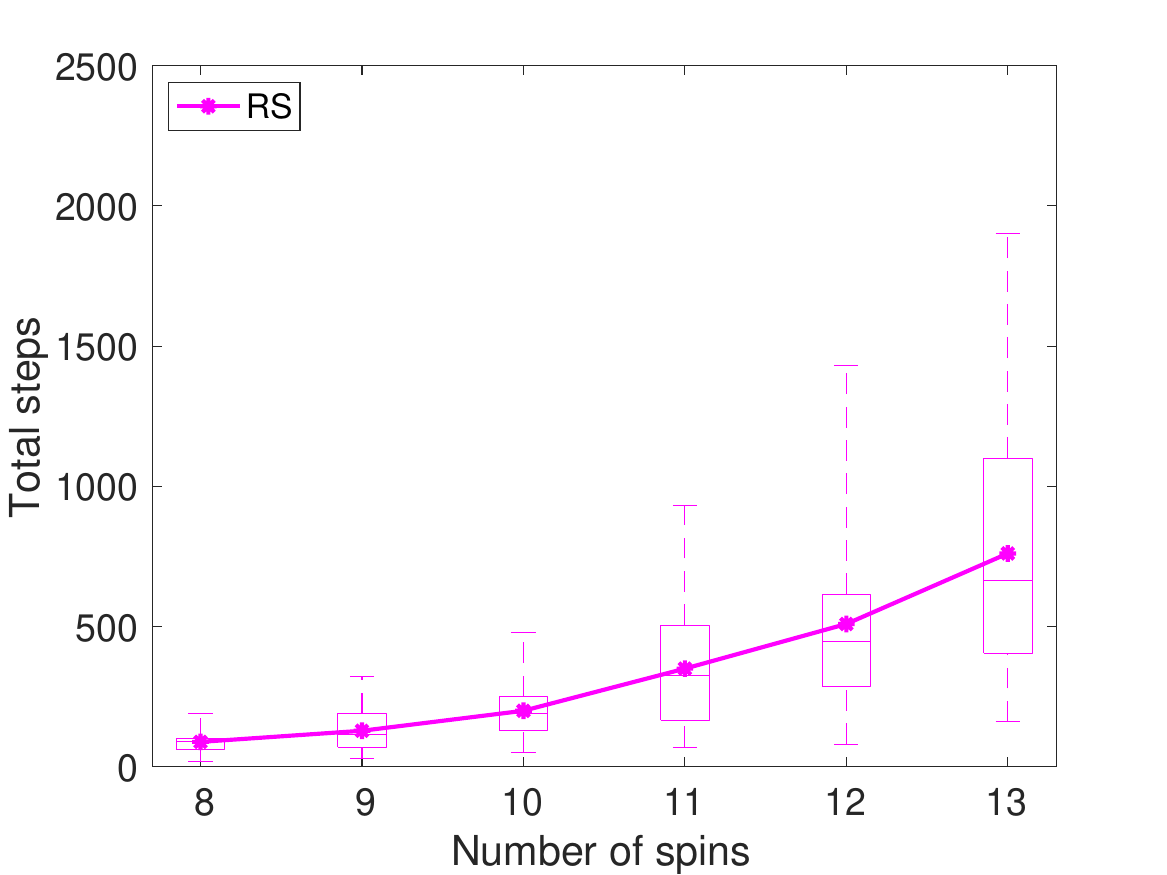} \\

    \subfigimg[width=\linewidth]{(c)}{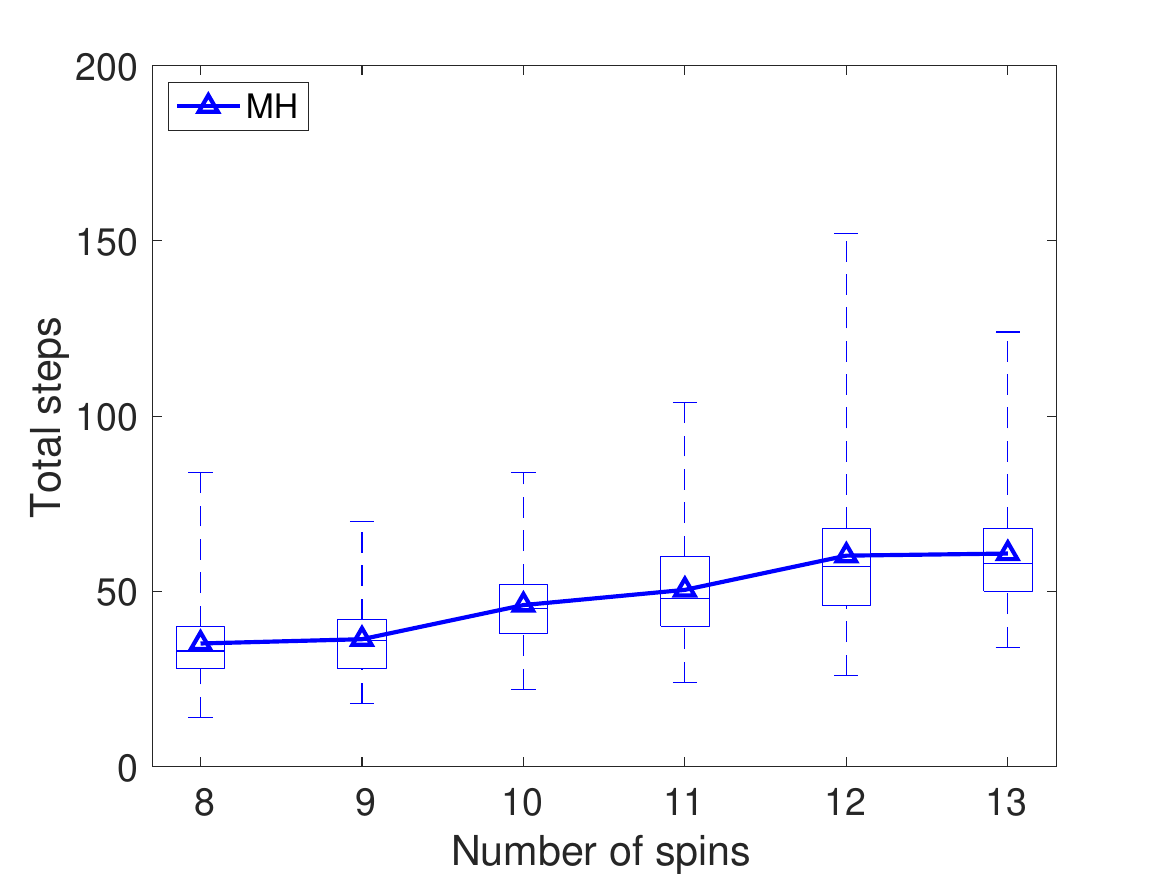} \\
    \end{tabular}
    \caption{ \textit{Box plots of the number of steps over $100$ randomly generated instances for Brute Force (BF), Random Sampling (RS), and Metropolis-Hastings (MH) algorithms.} Part (a) illustrates the steps required to achieve the optimal solution using BF, while part (b) shows the steps needed for the $0.9$-average energy approximation ratio using RS, and part (c) presents the results for the MH algorithm.}
    \label{fig:Box_Cl}
\end{figure}

\begin{figure}
    \centering
    \begin{tabular}{@{}p{0.7\linewidth}}
    \subfigimg[width=\linewidth]{(a)}{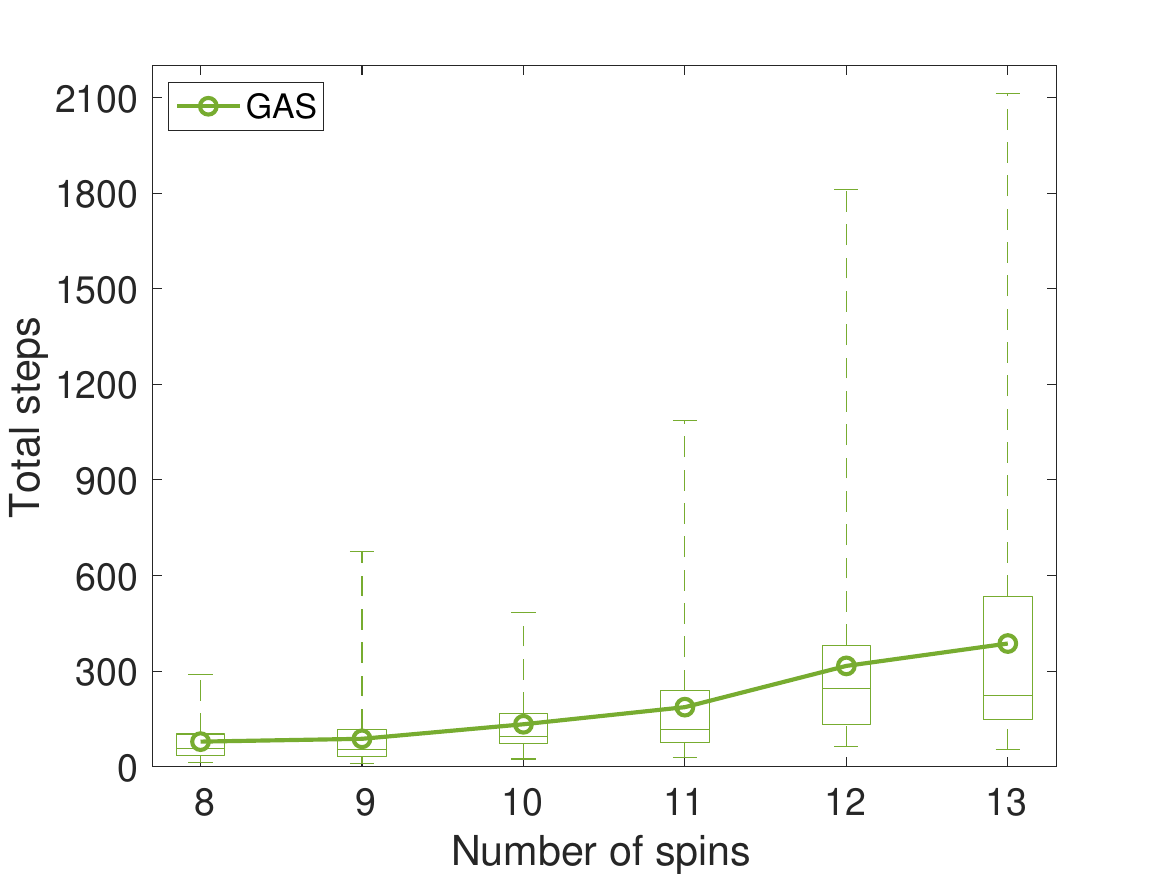} \\
    \subfigimg[width=\linewidth]{(b)}{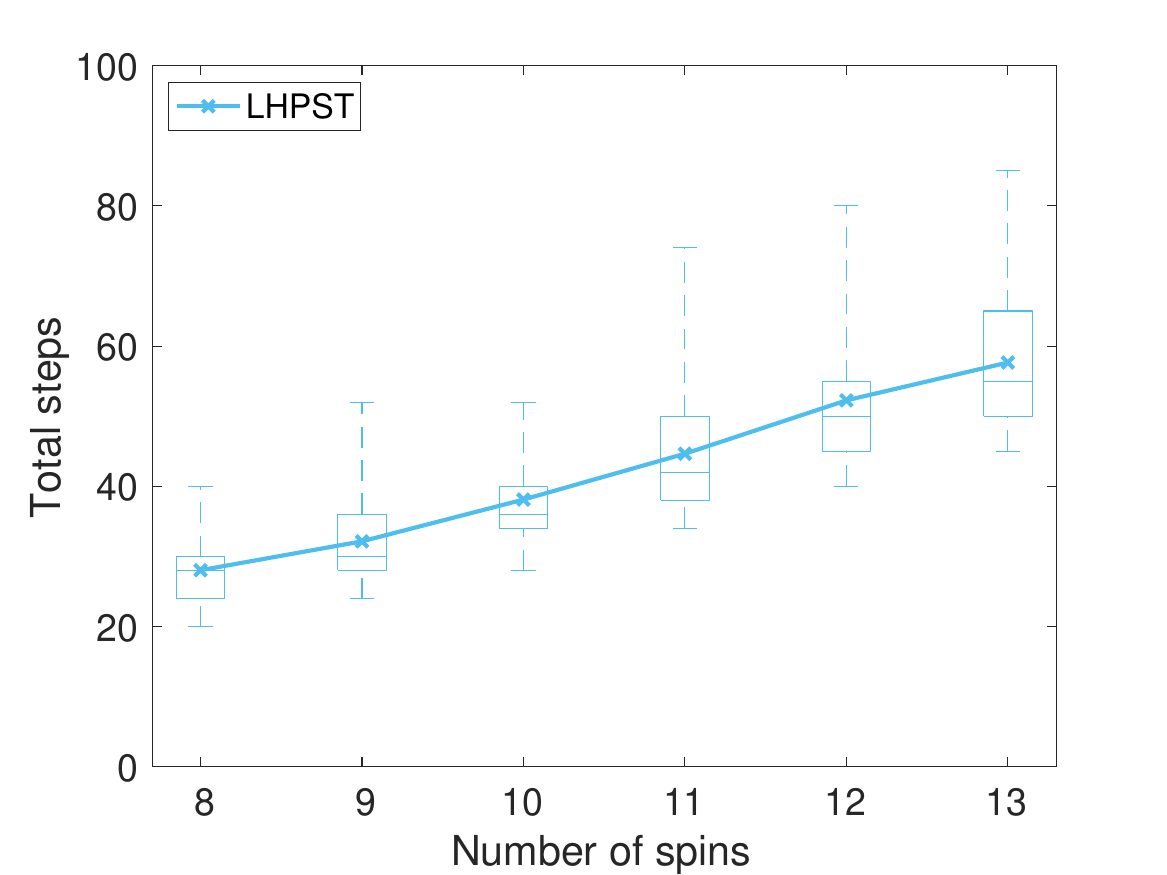} \\
    \end{tabular}
    \caption{\textit{Box plot results of the number of steps over $100$ randomly generated instances to obtain the optimal solution.} Part (a) illustrates the performance of Grover's algorithm, while part (b) depicts the steps required to achieve the $0.9$-average energy approximation ratio using the LHPST method.}
    \label{fig:Box_Q}
\end{figure}

This section includes the box plot for each of the algorithms used in \cref{fig:main_plot}, classical methods in \cref{fig:Box_Cl} and quantum methods in \cref{fig:Box_Q}. We also included in \cref{fig:ratio_plot} the \textit{ratio gap}, ratio between the first, $E_1$, and the ground state, $E_{gs}$, over the SK instances tested and beyond. 

\clearpage

\end{document}